\documentclass[prb,aps,amsmath,showpacs,preprint]{revtex4}

\usepackage{graphicx}

\usepackage{dcolumn}

\usepackage{bm}

\begin{document}

\title{First Principles Electronic Structure of Mn doped GaAs, GaP, and GaN semiconductors} 

\author{T. C. Schulthess$^{1}$, W. M. Temmerman$^{2}$, 
Z. Szotek$^{2}$, A. Svane$^{3}$, L. Petit$^{1}$} 
\affiliation{$^{1}$Computer Science and Mathematics Division and Center 
for Nanophase Materials Sciences, Oak Ridge National Laboratory, 
Oak Ridge, Tennesee 37831-6164, USA\\
$^{2}$Daresbury Laboratory, Daresbury, Warrington WA4 4AD, UK \\
$^{3}$Department of Physics and Astronomy, University of Aarhus, DK-8000 Aarhus C, Denmark
}
\date{\today}

\newpage

\begin{abstract}
We present first-principles electronic structure calculations of Mn doped III-V  semiconductors based on the local spin-density approximation (LSDA) as well as the self-interaction corrected local spin density method (SIC-LSD). We find that it is crucial to use a self-interaction free approach to properly describe the electronic ground state. The SIC-LSD calculations predict the proper electronic ground state configuration for Mn in GaAs, GaP, and GaN. Excellent quantitative agreement with experiment is found for magnetic moment and $p$-$d$ exchange in (GaMn)As. These results allow us to validate commonly used models for magnetic semiconductors. Furthermore, we discuss the delicate problem of extracting binding energies of localized levels from density functional theory calculations. We propose three approaches to take into account final state effects to estimate the binding energies of the Mn-$d$ levels in GaAs. We find good agreement between computed values and estimates from photoemisison experiments. 

\end{abstract}
\maketitle


\section{introduction}

Semiconductor spintronics aims to exploit both the spin and charge of electrons
in new generations of fast, low dissipation, non-volatile integrated information
storage and processing devices.
The Mn doped Ga-V semiconductors are amongst the most interesting materials
for applications in such new devices. In particular, (GaMn)As has been established 
as a well-behaved mean field ferromagnet with 
the Curie temperature T$_{c}$ linearly dependent on the concentration of the 
substitutional Mn,\cite{Jungwirthetal:PRB2005} a magnetic moment per Mn close to 
its free ion value,\cite{Edmonds:PRB05} and a possible low concentration of 
carriers 
further promoting T$_{c}$.\cite{Wangetal:JAP2004} Recent advances 
in thin-film growth of III-V semiconductors doped with Mn have led to synthesis 
of (GaMn)As dilute magnetic semiconductor (DMS) with 
Curie temperatures of the order of 173 K\cite{Foxon:JMatSci04}. 
Also, recent experiments by Edmonds et al.\cite{Edmondsetal:PRL2006} indicate
a carrier-induced nature of the ferromagnetic exchange, but a small, finite, density of 
unoccupied Mn $d$ states is found close to the Fermi level, reflecting hybridization 
with the host valence bands. Burch et al.,\cite{Burchetal:2006} on the other hand, claim 
observing impurity band conduction in Ga$_{1-x}$Mn$_{x}$As, with large effective 
masses of the carriers, so far not confirmed by other experiments. 

The most interesting characteristic of DMS is the carrier induced nature of the 
magnetic coupling, which has two important practical implications. First, 
the carriers are polarized and DMS can serve as efficient sources for spin 
injection, owing to the fact that they are structurally compatible with semiconductors 
used in devices, which alleviates the problems of interfacial disorder that prohibits 
efficient spin-injection from traditional ferromagnets into semiconductors. 
Second, because the Curie temperature is correlated with the carrier 
concentration, the magnetic order can be manipulated with voltage\cite{Ohno:Nature00}. 
Therefore, efforts aimed at increasing the Curie temperature of magnetic 
semiconductors have to be concerned with the nature of the magnetic exchange 
coupling in order not to loose these main advantages of carrier induced magnetism.
However, in order that these materials 
be relevant for practical
spin- and magneto-electronics\cite{Wolf:Science01}  applications,
their Curie temperatures have to be raised above room temperature. 

The successful description of magnetic properties in 
${\mathrm{Ga}}_{1-x}{\mathrm{Mn}}_{x}{\mathrm{As}}$ motivated Dietl 
et al.\cite{Dietl:Science00} to use the Zener model\cite{Zener:PR50,Zener:PR51b} 
description to predict Curie temperatures of various Mn doped group IV, III-V, 
and II-VI semiconductors. In particular, their prediction of a high Curie temperature 
of Mn doped GaN inspired many groups to synthesize this system,
 and several reports 
of Curie temperatures well in excess of room temperature now exist in the 
literature\cite{Sonoda:JCrystGrowth02}. 
However, despite important advances made for the Ga$_{1-x}$Mn$_{x}$As system, 
the microscopic nature of the electronic structure and magnetic exchange of 
Ga-V systems, and especially (GaMn)N, is far from understood. 
In particular, it still needs to be established whether the magnetism in these 
materials complies with the description provided by the Zener model. 
More fundamentally, it is not clear to what extent the Kondo-like Hamiltonian, 
describing Mn spins of $S$=5/2 
interacting with free carriers, is a justified starting point for describing 
systems other than Mn doped GaAs. For example, electron spin resonance 
measurements, which indeed support the 
picture of divalent Mn in GaAs\cite{Schneider},   
clearly favour the trivalent $d^{4}$ 
configuration for Mn impurities in GaP\cite{Kreissl}.
For GaN it has also been reported that Mn is in a divalent state when electrons are
doped\cite{Grafetal:PRB2003}, but in a trivalent state when holes are doped to
the system.\cite{Korotokovetal:PhysB2001} Hwang et al.,\cite{Hwangetal:PRB2005} 
using photoemission and soft X-ray absorption spectroscopy, confirm that in the 
$n$-type doped GaN the Mn state is divalent, while for the non-doped one it is 
trivalent.

Most model descriptions of Mn doped GaAs assume the Mn impurity to have a localized 
moment of $S$ = 5/2, formed by five occupied atomic-like $d$-orbitals ($d^{5}$) 
that interact weakly with the host valence band through level repulsion,
leading to the simple ($d^{5}+h$) picture where the Mn spin couples antiferromagnetically to a 
 polarized hole, $h$. The resulting Kondo-like Hamiltonian appears well justified 
with parameters that can be established experimentally. However, its solution is 
still under debate. Dietl et al.\cite{Dietl:PRB01,Dietl:Science00},  as well as Mac 
Donald and co-workers\cite{MacDonald:Review02}  use the Zener\cite{Zener:PR50,Zener:PR51b}  
description, in which it is assumed that the ferromagnetic interaction between Mn ions 
is mediated by the induced spin-density that is anti-aligned with the Mn moments, and 
the resulting prediction of the Curie temperature, $T_{c}$, appears in agreement with 
experiment for ${\mathrm{Ga}}_{1-x}{\mathrm{Mn}}_{x}{\mathrm{As}}$, with $x$ up to 
0.09\cite{Foxon:JMatSci04}. 

The key question that needs to be addressed from an 
electronic structure point of view is whether at concentrations of several 
atomic percent of substitutional Mn, the acceptor level forms an impurity band 
broad enough to merge with the valence band.
The electronic structure of Mn in Ga-V 
semiconductor hosts is best studied with first principles electronic 
structure calculations. Since the Local (Spin-)Density Approximation\cite{vonBarth:JPhysC72}  
(L\{S\}DA) to Density Functional Theory\cite{Hohenberg:PR64,Kohn:PR65}  (DFT) has proven 
very successful in predicting ground state properties in semiconductors\cite{martin}  
as well as transition metals and their alloys \cite{kubler}, several 
groups~\cite{Sato:JJAP2001,Schulthess:JAP01,Sato:Review02,Mahadevan:PRB04,Mahadevan:APL04,Filippetti:JMMM2005} 
have performed 
LSDA calculations of the electronic structure and magnetism of various transition metal 
doped semiconductor systems soon after the discovery of ferromagnetism in Mn doped GaAs. These studies all agree in one important 
point\cite{Mahadevan:PRB04}: the Mn $d$-orbitals are not atomic-like but hybridize 
rather strongly with the host valence band (As-$p$ in the case of 
${\mathrm{Ga}}_{1-x}{\mathrm{Mn}}_{x}{\mathrm{As}}$). In this, the LSDA picture 
differs substantially from the one of model calculations. We have recently 
shown \cite{Schulthess:NMAT05}, that first principles calculations based on the 
self-interaction corrected local spin density (SIC-LSD) method, lead to a picture that is 
more consistent with what is expected
for a system with strongly correlated electrons. 
The purpose of the present paper is to make a more in-depth comparison between the 
electronic structure results using different functionals and comparing the first principles predictions with experiments. We begin with a discussion of the SIC-LSD method, how it is applied in Mn doped III-V systems and how it compares to LDA and other methods used to "fix" LDA with respect to electron correlations. Then we compare the results of the SIC-LSD calculations for Mn doped GaAs with experiment and previously known LDA results. Finally, by studying the changes in electronic configuration and trends for different III-V systems, we shed some light on the intricacy of the electronic structure of Mn doped GaAs.

\section{Spurious self-interactions and their removal}
The local (spin) density approximation 
to density functional theory 
forms the foundation of the conventional band theory. It describes electron correlations
at the level of the homogeneous interacting electron gas and has been very successful in
predicting electronic properties of many materials in terms of their ground state charge 
density.\cite{jones,kuebler} For semiconductors, this is particularly true for the prediction of structural properties and alloying behavior. A common criticism of LDA and its generalization, the generalized gradient approximation (GGA), namely the systematic tendency to underestimate the band gap, is strictly speaking not a shortcoming, since DFT in its present use is a theory of the ground state and hence not meant to predict quasi particle excitation spectra. Nevertheless, LDA and GGA based band structures are often used to describe the valence and conduction bands in semiconductors and numerous methods have been applied to correct the magnitude of the band gap.\cite{SSP:2000}

In solids with localized $d$- and/or $f$-electrons, such as the transition metal monoxides, the cuprate high temperature superconductors, rare earths, actinides, and, as we will see below, dilute magnetic semiconductors, LDA notoriously fails to describe correctly the electronic and magnetic ground state properties. This can be understood as follows: in Kohn-Sham based DFT, the energy functional of the electron density is written as,

\begin{equation}
E[n] = T_s[n]  + \int{V_{\mathrm{ion}}(r) n(r) dr} + J[n]+E_{xc}[n],
\end{equation}
where $T_s$ is the single particle kinetic energy, $V_{\mathrm{ion}}$ the ionic potential, $J$ the classical Coulomb interaction energy - or Hartree energy - of the charge density, and $E_{xc}$ is the exchange and correlation energy that contains the exchange term and all the non-classical electron correlations. When the electron density is decomposed into orbital densities, $n=\sum_i{n_i}$, it is straightforward to demonstrate that the Hartree term contains a contribution, $J[n_i]$, of an orbital interacting with itself. This self-interaction term is cancelled exactly by the self-exchange contribution to $E_{xc}$. In the LDA and GGA, $E_{xc}$ is approximated and the self-interactions are not cancelled anymore. One speaks of spurious self-interactions that are introduced by the approximations in the LDA and GGA functionals.\cite{DreizlerGross}

The spurious self-interactions are negligible for extended orbitals such as the $s$ and $p$ bands in semiconductors or $d$ bands in transition metals. They are, however, substantial whenever electrons occupy localized orbitals such as the  3$d$ orbitals of transition metal atoms in oxides or of transition metal impurities in semiconductors. In these cases, the spurious self-interactions push the localized orbitals into the valence band usually resulting in too strong a hybridization with the other valence electrons.

This problem was recognized already more than two decades ago and a remedy was proposed by Perdew and Zunger \cite{PZ:SIC} to simply subtract the spurious self-interactions from the LSDA functional. Their self-interaction corrected (SIC) local spin density (LSD) functional takes the form

\begin{equation}
E^{\mathrm{SIC}} = E^{\mathrm{LSDA}} - \sum_{i}^{\mathrm{occ.}}\delta_{i}^{\mathrm{SIC}}
\label{SIC-LSD}
\end{equation}
where the sum runs over occupied and localized orbitals with non-vanishing self-interaction corrections

\begin{equation}
\delta_{i}^{\mathrm{SIC}} = J[n_i] + E_{xc}^{\mathrm{LSDA}}[n_i].
\label{SIC}
\end{equation}

When applied to atoms, the most extreme case where all electrons occupy localized orbitals, the SIC-LSD functional drastically improves the description of the electronic structure \cite{Norman:PRB84}. In solids, where not all electrons occupy localized orbitals, one is faced with the task of minimizing the {\it orbital dependent} SIC-LSD functional (\ref{SIC-LSD}). Additionally,  since the LSDA exchange correlation functional depends non-linearly on the density, the self-interaction corrections (\ref{SIC}) and hence the SIC-LSD functional (\ref{SIC-LSD}) are {\it not} invariant under unitary transformations of the basis and one is thus faced with a daunting functional minimization problem.

Since the main effect of the self-interaction correction is to reduce the hybridization of localized electrons with the valence band, the technical difficulties of minimizing the SIC-LSD functional in solids can often be circumvented by introducing an empirical Coulomb interaction parameter $U$ on the orbitals that are meant to be localized. The original derivation of the LDA+U approach\cite{Anisimov:PRB91} seems to have been based on the conjecture that LDA can be viewed as a homogeneous solution of the Hartree-Fock equations with equal, averaged, occupations of localized $d$- and/or $f$- orbitals in a solid. Therefore, as such, it can be modified to take into account the on-site Coulomb interaction, $U$, for those orbitals to provide a better description of their localization.  The on-site Hubbard $U$ is usually treated as an adjustable parameter, and chosen to optimize agreement with experiment. In many transition metal oxides, the approach is successful because the results are not very sensitive to the precise value of $U$. However, in Mn doped III-V systems, the opposite seems to be the case,\cite{Sandratskiietal:PRB2004} magnetic exchange and moment depend sensitively on the value of $U$ and the method looses its predictive power.

For this reason we are using an approach\cite{Temmerman:SIC} with which the full SIC-LSD functional is minimized with respect to the orbital decomposed charge density, giving rise to a generalized eigenvalue problem with an orbital dependent potential - as the self-interaction corrections are only non-zero for localized electrons \cite{PZ:SIC}, the localized and delocalized electrons experience different potentials. The latter move in the LSD potentials, defined by the ground state charge density of all occupied states, while the former experience a potential from which the self-interaction term has been subtracted. Hence, in this formulation one distinguishes between localized and itinerant states and it is possible to study different nominal valences for those elements in the solid that contain localized $d-$ and/or $f$-electrons. To determine the ground state energy and valence, one minimizes the SIC-LSD functional with respect to these electronic configurations (different distributions of localized and itinerant states). The resulting SIC-LSD method is a first principles theory for the ground state with no adjustable parameters. Finally, it is important to note that the SIC-LSD functional subsumes the LSDA, that is, when all electrons (besides the core electrons) are itinerant, the configuration in which no orbitals self-interaction correct will have the lowest energy and the solutions of the SIC-LSD functional will be identical to that of the LSDA.

\section{Electronic structure of GaMnAs}

In this study we model the idealized \cite{footnote}
Mn doped semiconductor with a supercell approach with a unit cell of up to 64 atoms, in which one Ga atom has been 
replaced with a Mn impurity. We limit ourselves to the zinc-blende structure with experimental lattice constants of the host semiconductor and neglect structural relaxations. In all calculations, the 4$s$ and 4$p$ electrons of the semiconductor hosts are itinerant, forming the valence band - applying self-interaction corrections to these electrons raises the energy. Self-interaction corrections are applied to states that originate from the Mn-3$d$ electrons, that is, we minimize (\ref{SIC-LSD}) over orbital decomposed densities and the different combinations of applying self-interaction corrections to the Mn-3$d$ levels. Of all possible combinations, the following two scenarios always have the lowest energy: (1) when self-interaction corrections are applied to all five majority Mn-3$d$ levels - this is the $S=5/2$ configuration which we denote by Mn($d^5$) or Mn$^{2+}$; (2) when self-interaction corrections are applied to all but one of the majority Mn t$_{\mathrm{2g}}$ levels - this configuration has $S=2$ and we will call it Mn($d^4$) or Mn$^{3+}$.

\subsection{Ground state, magnetic moment, and exchange}

The relative energies of the three relevant configurations for one Mn impurity in a 64 atom unit cell of GaAs are $E(d^4)-E(d^5)=0.18$ eV and $E(LSDA)-E(d^5)=1.99$ eV. The lowest energy state (ground state) of this system is reached when all five majority $d$ electrons localize on the Mn atoms - the same is true for higher Mn concentrations (smaller unit cells). The ground state predicted by these calculations agrees with electronic configuration inferred from electron spin resonance measurements \cite{Schneider} . Both the $d^5$ and the $d^4$ scenarios are favourable when compared to the LSDA solution (for which none of the five Mn $d$ orbitals localize). The magnetic moment on the atomic sphere surrounding the Mn atoms for the $d^5$, $d^4$, and LSDA scenarios is 4.50$\mu_{\mathrm{B}}$, 4.07$\mu_{\mathrm{B}}$, and 3.80$\mu_{\mathrm{B}}$ respectively. The value for the ground state is in excellent agreement with the local Mn moment measured experimentally using X-ray magnetic circular dichroism (XMCD).\cite{Edmonds:PRB05} 

\begin{figure}[h]
\includegraphics[clip,scale=1.0,angle=0]{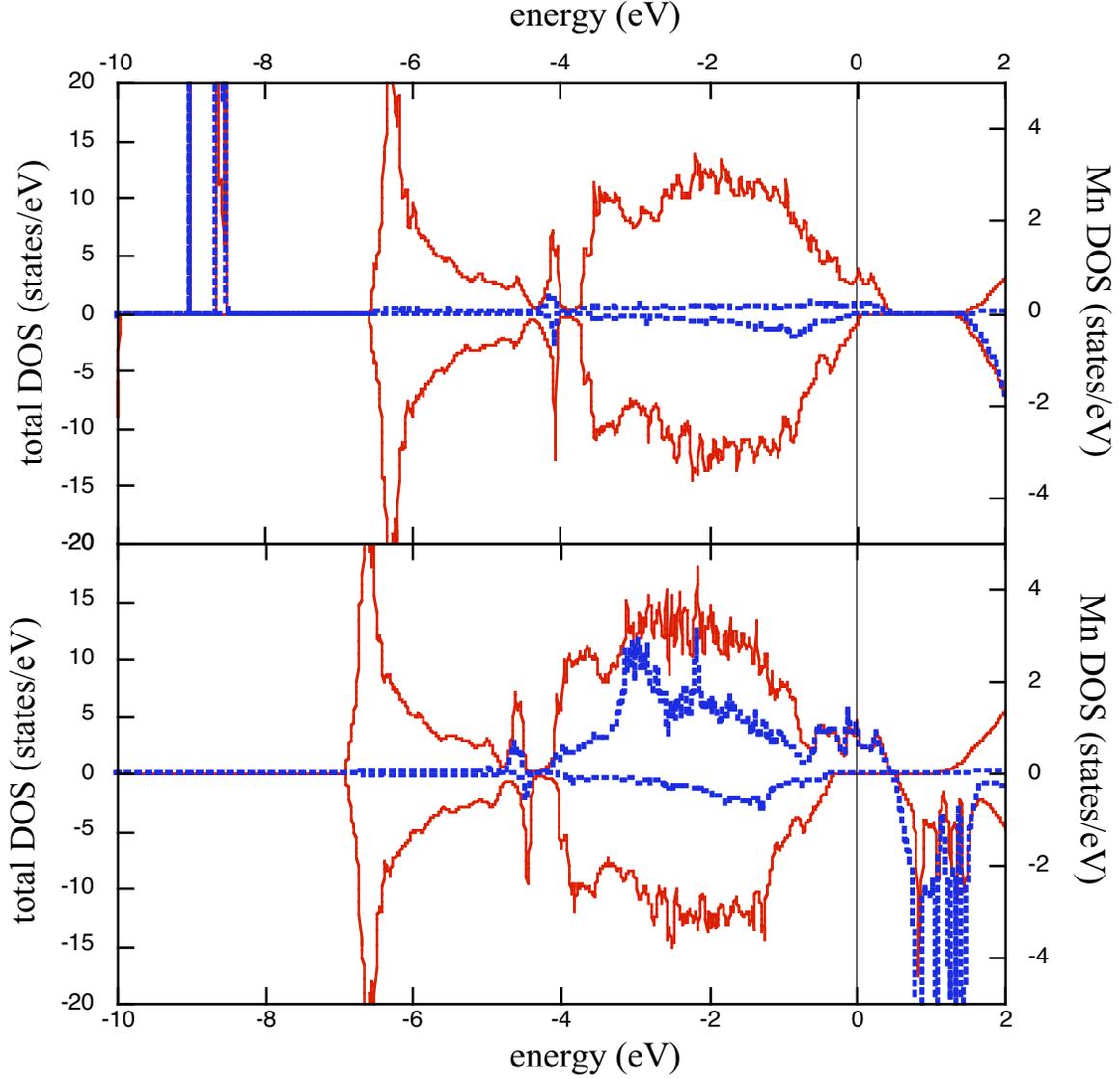}
\caption{
The spin-resolved total DOS (solid red line) and  Mn partial DOS (dotted blue line) of (Ga$_{1-x}$Mn$_x$)As with $x=6.25\%$ for the SIC-LSD ground state, $d^5$, scenario (upper panel) as well as the LSDA (lower panel). 
\label{gamnas_dos}
}
\end{figure}

A detailed characterization of the electronic state of Mn impurities in GaAs and the 
magnetic coupling can best be seen from the density of states (DOS), 
which is plotted in figure \ref{gamnas_dos}. In the $d^5$ ground state scenario, all five 
majority $d$-states form a localized impurity band below the host valence band with virtually no hybridization. However, because of repulsion between the majority Mn-$d$ levels and the As-$p$ bands, the top of the host valence band is spin split, leading to a spin polarized hole. In this scenario, the hole has primarily As-$p$ character. This is a first principles electronic structure description of what is called the ($d^5+h$) state in the models, where the Mn impurity with $S=5/2$ is surrounded by a hydrogenic hole of opposite spin polarization. The spin polarization of the hole is, however, not 100\%, since, as can be seen from the plot, the Fermi level crosses both the majority and the minority bands. In other words, the spin splitting $\Delta$ of the valence band is smaller than the difference, $E_{\mathrm{VBM}}-E_{\mathrm{F}}$, between  the valence band maximum and the Fermi level.
That the hole is not strongly coupled to the Mn impurity, can also be inferred from its spacial extent: in our calculations the induced hole in the valence band extends beyond the size of the unit cell. Hence, contrary to what 
has been assumed in some models of Mn doped GaAs, the coupling between the Mn $S=5/2$ spin 
and the host valence band is not strong, and the Zener model based approach\cite{Dietl:Science00,MacDonald:Review02} to estimating the Curie temperature is justified. From the splitting of the valence band maximum we can estimate the $p-d$ exchange coupling (see ref. \cite{Schulthess:NMAT05} for details), the predicted values are given in Table \ref{jpd}. The agreement with the experimental value extracted from spin polarized photoemission data \cite{fujimori:PRB1999,fujimori:PRB2002} is very good. The ($d^5+h$) picture for Mn doped GaAs seems to be well justified by the first
 principles calculations. However, this applies        only in the concentration range below ~10 at. \% Mn. 
In calculations with 8
-atom unit cells, corresponding to Mn concentrations of 25 
at. \%, respectively, the valence band 
is strongly perturbed by the Mn alloying effect and the $p-d$ exchange parameter extracted from 
the valence band splitting is concentration dependent. The simple Zener model is probably not 
valid at higher Mn concentrations and one should not simply extrapolate the Curie temperature 
predictions of this model to concentrations higher than 10 at. \% Mn.

The effects of the spurious self-interactions in the LSDA are most clear in the density of states \ref{gamnas_dos}. Since localizing the Mn-$d$ orbitals is energetically unfavourable due to the unphysical self-interactions, the $d$ levels are pushed up into the valence band and hence become strongly hybridized. As a result, the local Mn moment (3.8 $\mu_{\mathrm{B}}$ for LSDA) is underestimated. But more importantly, the nature of the states at the Fermi level is very different leading to a different mechanism for the magnetic coupling. The $p-d$ exchange coupling is strongly overstated, the splitting of the top of the valence band being a factor of two larger than experiment (Table \ref{jpd}.

\begin{table}[t]
\caption{Spin splitting at the top of the valence band, $\Delta$ and the $p-d$ exchange coupling parameter $J_{pd}$ and the more commonly quoted $N_0 \beta$ parameter ($N_0$ is the number of unit cells in the normalization volume and $\beta = -J_{pd}$). We are comparing the ground states of the SIC-LSD and the LSDA functionals with experiment.}
\begin{tabular}{|c|c|c|c|c|} 
\hline
  & $x$ & $\Delta$ & $J_{pd}$ & $N_0 \beta$\\
  &         & (eV) & (meV nm$^3$) & (eV)\\
   \hline
 Mn($d^5$) & 0.25 & 0.69 & 27 & -0.60 \\
  & 1/16 & 0.43 & 68 & -1.51 \\
  & 1/32 & 0.21 & 67 & -1.49 \\
\hline
LSDA & 1/16 & 0.71 & 114 & -2.5 \\
            & 1/32 & 0.45 & 145 & -3.2 \\
\hline
Experiment &   &   & 54$\pm$9 & -1.2$\pm$0.2 \\
 \hline
\end{tabular}
\label{jpd}
\end{table}

\subsection{ Localized levels and photoemission}

Although DFT is strictly speaking a theory of the ground state from which spectroscopic information 
is not easily extracted, the LDA based band-structure is often compared to photoemission experiments. This is because the effective Kohn-Sham potentials can be viewed as an energy independent self-energy and hence the Kohn-Sham energy bands correspond to the mean field approximation for the spectral function. In the SIC-LSD, this argument only applies to the itinerant orbitals that are not self-interaction corrected. The localized states that have been self-interaction corrected see
a different potential\cite{Temmerman:SIC}, and the solution (or the solutions) 
to the generalized SIC-LSD eigenvalue problem, which is different from the solution to
 the Kohn-Sham equations in the LDA, no-longer correspond to a mean field approximation of the spectral function. To extract spectroscopic information for the localized Mn $3d$ levels from the SIC-LSD calculations, we have to take a different route. We consider three approaches: 

1. The canonical approach to compute photoemission energies of localized states in the literature is a $\Delta_{\mathrm{SCF}}$ calculation \cite{Freeman:87}, in which the total energy differences between two configurations of Mn assumed to represent the initial and final states of a photoemission experiment are compared. This is expressed by the formula 
\begin{equation}
\Delta{^{(1)}}_{\mathrm{SCF}} = E(d^{4}, N_{pd} + 1) - E(d^{5}, N_{pd}),
\end{equation}
meaning that one removes an electron from a Mn $d$ level and introduces it at the top of the valence
band, thus increasing the total number of valence electrons in the supercell, $N_{pd}$, by 1. Here $E$ represents the
total energies of the respective $d^{4}$ and $d^{5}$ configurations. For $E(d^4,N_{pd}+1)$ the electronic configuration is constrained such that the fifth $d$ level is unoccupied.
Of course, a problem one faces with this kind of estimate is that the correct final state 
might not be the ground state of the
system with one electron removed. For example, if a final state of $d^4$ is assumed for Mn in GaAs,
and the constraint on the fifth $d$ level is dropped during iterations to self-consistency it will move below the  valence band maximum (VBM), and the level starts to get filled (until eventually the filling balances the position of the level at the Fermi energy). Hence, in the end the final state corresponds to some 
intermediate Mn configuration between $d^{5}$ and constrained $d^{4}$. The effect is partially physical, since the screening processes
around a hole created by photoexcitation include Mn $d$ screening, however not by the
state being kicked out but by all the minority $d$ electrons which become attracted to the 
$d$-hole. An alternative interpretation of the photoemission experiment is given by the formula
\begin{equation}
\Delta{^{(2)}}_{\mathrm{SCF}} = E(d^{4}, N_{pd}) - E(d^{5}, N_{pd}-1).
\end{equation}
where one simply compares two final state energies: the energy of the system with one Mn $d$ electron removed compared to the energy of the system with one electron removed at the Fermi level. This alternative way of interpreting the photoemission spectrum leads to the numerical result that is very similar to $\Delta^{(1)}_{\mathrm{SCF}}$, as can be seen from Table \ref{d_levels}.

\begin{table}[t]
\caption{Positions of the Mn 3$d$ level (in eV) as they would appear in photoemission experiments computed with the methods discussed in the text.}
\begin{tabular}{|c|c|c|c|c|} 
\hline
 $\Delta^{(1)}_{\mathrm{SCF}}$ & $\Delta^{(2)}_{\mathrm{SCF}}$ & OEP & $\epsilon_{\mathrm{TS}}$ & experiment \cite{fujimori:PRB1999}\\
 \hline
 3.23 & 3.16 & 3.7 & 5.1 & 4.2 \\
 \hline
\end{tabular}
\label{d_levels}
\end{table}


2. In the earlier section we have described the SIC-LSD method as an orbital dependent 
density functional theory. Formally, however, SIC-LSD may be viewed as a standard 
density functional theory, implying that the SIC-LSD 
energy functional can be represented as a functional of the total charge density alone
and minimized with respect to it.
This means that there exists an effective Kohn-Sham equation with an effective potential,
which is common to all Kohn-Sham states, is self-interaction free, and
depends only on the total charge density. Here the situation is completely analogous to the 
optimized effective potential (OEP) introduced in connection with the Hatre-Fock 
approximation,\cite{iafrate:PRA92,iafrate1:PRA92,iafrate:PRA93,kotani:JPCM98} for which 
case the Kohn-Sham eigenvalues are often
compared to quasiparticle energies, with a considerable improvement over the Hartree-Fock
eigenenergies. 

\begin{figure}[h!]
\includegraphics[scale=1.0,angle=0]{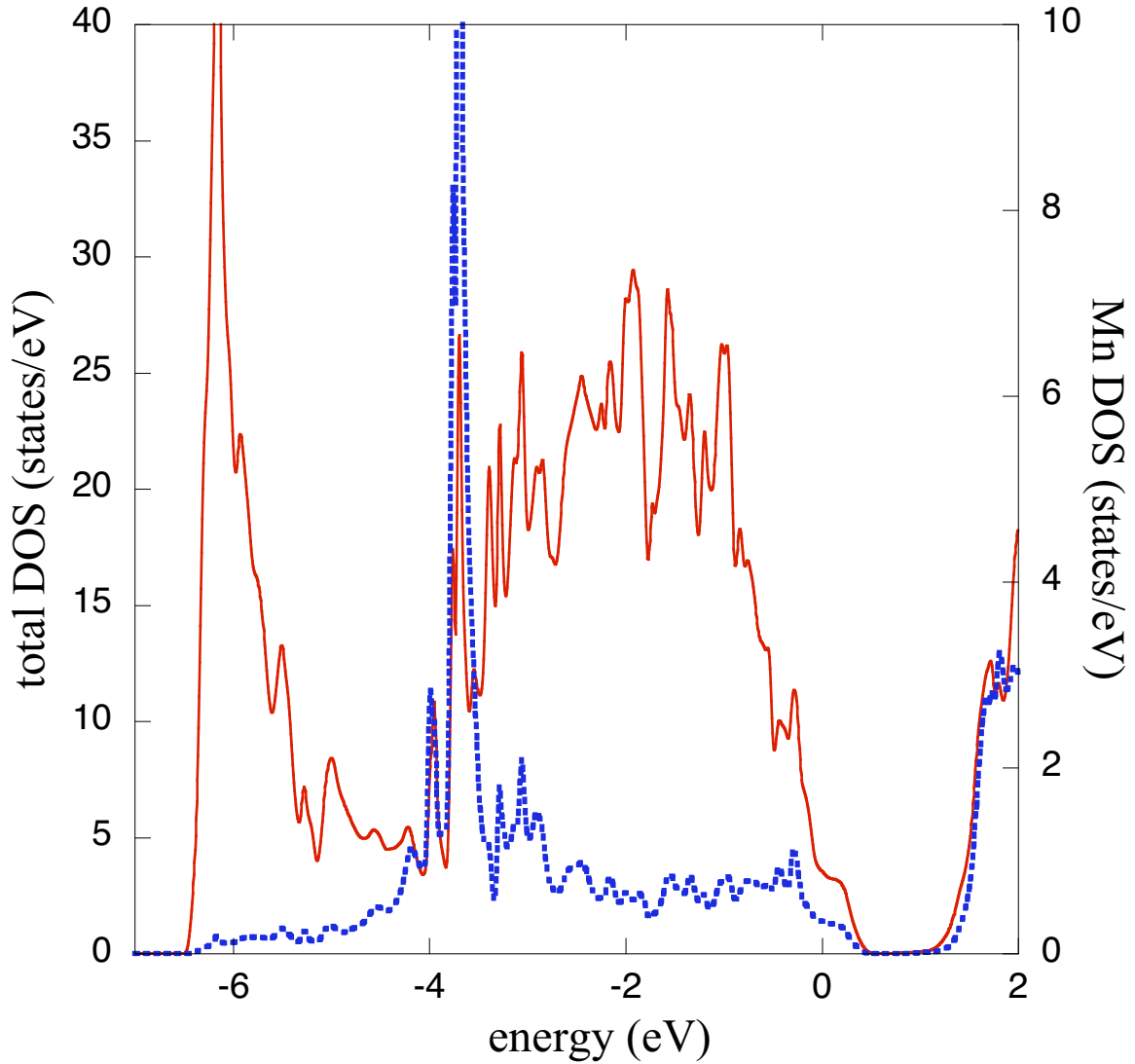}
\caption{Density of states from SIC-based OEP calculations, showing both the total (solid red line)
and Mn $d$ (dotted blue line) contributions.
\label{OEP_Mn}
}
\end{figure}

Adopting the OEP philosophy, one can search for the effective potential, which reproduces 
the SIC spin-density, and with such a potential derive the density of states of Mn in GaAs. 
Since the only significant
difference between the LSD and SIC-LSD charge densities for Mn in 
semiconducting hosts is the Mn spin moment, we can constrain the search by looking for 
that particular potential shift on the Mn site, which will reproduce the self-consistent spin 
moment of Mn in the SIC-LSD calculations. 
Hence, as in the standard OEP approach, the derived eigenvalues
reflect the self-interaction correction in that the Mn majority $d$-states lie lower in
energy than in the LSD case, however not as low as when calculated directly from the SIC eigenvalues.
The OEP-like SIC-derived density of states is shown in figure \ref{OEP_Mn}. One can see nearly
perfect agreement with the experimentally determined position of the Mn-$d$ states (Table \ref{d_levels}).

3. As a 'quick and dirty' way to obtain electron removal energies just from the self-consistent
ground state calculation one may apply a transition state approximation, according to which
the Mn removal energy is defined as the average of the 
SIC-LSD and LSD $d$ state positions:
\begin{equation}
\epsilon_{\mathrm{TS}}^-=\frac{1}{2}(<f|H_{\mathrm{LSD}}+V_{SIC}|f>+<f|H_{\mathrm{LSD}}|f>).
\label{epsTS2}
\end{equation}
In effect, the SIC potential is only counted with half of its strength in the transition state approximation to
the removal energy.
By evaluating $H_{LSD}$ in the initial state, {\it i.e.} without the hole in the $d$ shell,
 we avoid the aforementioned 
effect of the $d$-hole pulling the $d$ levels down. 
This approximation has been discussed for rare-earth impurities in GaAs and GaN.\cite{DMSrareearths}
The transition state philosophy was also implemented in
Ref. \onlinecite{Filippetti:PRB2003}, albeit in a different manner, by invoking the averaging factor of 
$\frac{1}{2}$ already
in the total energy functional, while we do it here only for the removal energy, Eq. (\ref{epsTS2}), 
after self-consistency. 

In Table \ref{d_levels} we compare the three methods just described to estimate the positions of the $d$ levels in photoemission experiment of Mn doped GaAs. The agreement with experiment is satisfactory, but certainly not as perfect as the quantitative agreement we find for the magnetic moment (discussed above). This should, however, come as no surprise, since DFT is a theory for the ground state and the magnetic moments that are determined from the spin densities are rigorously founded in spin depended DFT. Similar results for the $d$ levels have been achieved with the LDA+U method by Shick et al.,\cite{Shicketal:PRB2004} assuming 
$U$ of 4 eV. Note, however, that in that study they report a large sensitivity of the Mn $d$ binding 
energies to the magnitude of $U$, which is not the case in the present work, as there is no adjustable 
parameter in the SIC-LSD method.

\section{Materials trends: Mn in GaAs, GaP, and GaN}

\begin{figure}[h]
\includegraphics[clip,scale=0.9,angle=0]{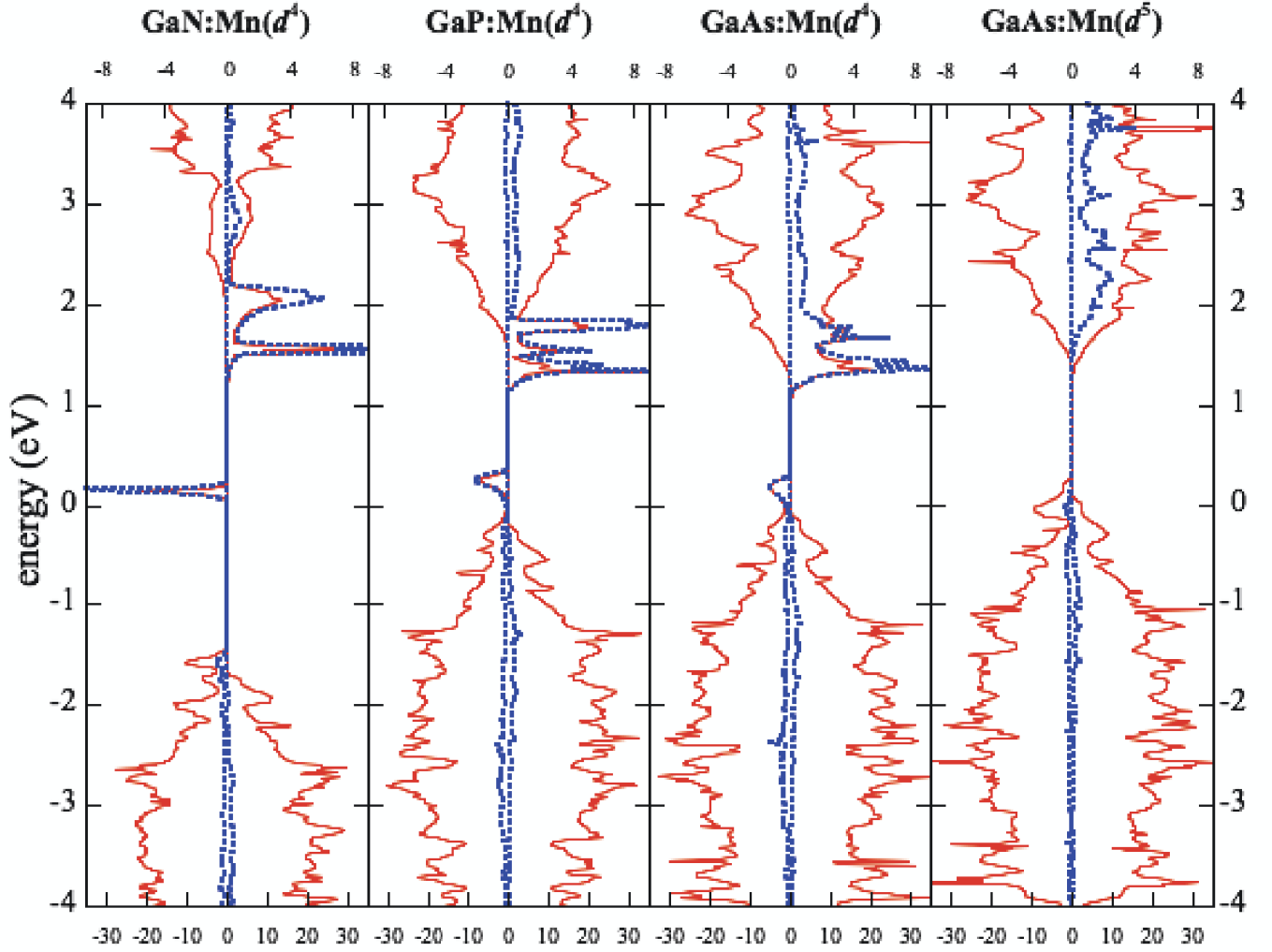}
\caption{
The spin-resolved total DOS (solid red line and bottom scale) and Mn-DOS (dotted blue line and top scale) from left to right, for Mn-$d^4$ impurity in GaN, GaP, and GaAs, as well as Mn-$d^5$ impurity in GaAs. Results are shown for 64 atom supercell (3.125\% Mn).
\label{III-V_DOS}
}
\end{figure}

The energy differences, $E(d^4)-E(d^5)$, between the Mn($d^4$) and Mn($d^5$) scenarios in a 64 atom unit cell of GaAs, GaP, and GaN are, respectively, 0.18 eV, -0.05 eV, and -1.28 eV. Hence, the ground state configuration of Mn changes from $d^5$ in GaAs to $d^4$ in GaP and GaN. While the energy difference between the configurations is very small for Mn doped GaP, the predicted $d^4$ ground state seems to be in agreement with electron spin resonance experiments \cite{Kreissl}. This change in valence of the Mn impurity can best be understood in terms of a localization/delocalization transition of the fifth Mn $d$ orbital in the majority channel. 
In figure \ref{III-V_DOS} we compare the density of states of the two scenarios of Mn in all three Ga-V systems studied here. In GaN, a charge transfer insulator, the fifth Mn $d$ level is unoccupied, forming a deep impurity level in the middle of the band gap, and there is virtually no hybridization with the valence band. In GaP, the lattice constant is increased, the band gap reduced, the valence band moves up and closer to the Mn $d$ impurity level, and hybridization between the two is already substantial. As the hybridization in the $d^4$ scenario further increases in GaAs, the occupation of the fifth Mn $d$ level has increased to the point, where, due to the larger lattice constant, it is now energetically more favorable to localize, leading to a $d^5$ ground state configuration for the Mn impurity and a hole with As $p$ character.

Since the SIC-LSD method is a static theory for the ground state, our present calculations do not account for fluctuations that are almost certainly playing an important role whenever two or more solutions are close in energy, i.e. close to the transition between localized and delocalized states. Quantum fluctuations as well as fluctuations at finite temperature may well lead to some admixture of Mn $d^4$ and Mn $d^5$ in GaAs and GaP. Furthermore, co-doping with additional donors or acceptors will likely also have an effect on the localization nature of the fifth Mn $d$ orbital.\cite{Petit:PRB2006} Additionally, since interstitial Mn impurities, which we did not consider here, give rise to a localized impurity band near the Fermi 
level \cite{Ernst:PRL2005}, the measured electronic structure in Mn doped GaAs can be affected by many factors that are sample dependent.

\section{Summary}

As we have seen, the use of standard band structure techniques, such as the LSDA, does not lead to satisfactory description of the electronic structure of Mn doped semiconductors. This is because spurious self-interactions introduced by the approximation to the exchange correlation functional lead to an incorrect description of the Mn $d$ orbitals as well as of the hole mediated magnetic exchange. Hence, to properly describe magnetic semiconductors, methods have to be used that do not suffer from spurious self-interactions.

Here, we have applied the SIC-LSD method to study the electronic structure and magnetic properties of Mn doped  III-V semiconductors. For (GaMn)As our calculations predict the correct electronic ground state configuration for Mn, with its magnetic moment of 4.5 $\mu_{B}$ and $p$-$d$ exchange in excellent agreement with experiment. In agreement with previously assumed models, we find that the Mn spin is weakly antiferromagnetically coupled to the hole that mediates the ferromagnetic exchange. By taking into account screening/relaxation effects, we have obtained a very good agreement of the calculated binding energies for the Mn $d$ states with photoemission experiments. Our calculations also describe correctly the trends of the electronic ground state for Mn in GaN, GaP, and GaAs. In particular we predict the localization/delocalization transition from Mn (d$^{5}$) to Mn (d$^{4}$), when moving from GaAs to GaP and GaN. As the Mn($d^5$) and Mn($d^4$) configuration are energetically very close in GaP and GaAs, it is likely that effects of co-doping, additional impurities, and thermal fluctuations lead to mixed valence states for the measured electronic ground state configuration of Mn that might be sample dependent.

\begin{acknowledgments}
This research used resources of the National Center for Computational Sciences at Oak Ridge National Laboratory. It was supported by the Division of Scientific User Facilities as well as the Division for Materials Science and Engineering of the U. S. Department of Energy.
\end{acknowledgments}


\begin{thebibliography}{54}
\expandafter\ifx\csname natexlab\endcsname\relax\def\natexlab#1{#1}\fi
\expandafter\ifx\csname bibnamefont\endcsname\relax
  \def\bibnamefont#1{#1}\fi
\expandafter\ifx\csname bibfnamefont\endcsname\relax
  \def\bibfnamefont#1{#1}\fi
\expandafter\ifx\csname citenamefont\endcsname\relax
  \def\citenamefont#1{#1}\fi
\expandafter\ifx\csname url\endcsname\relax
  \def\url#1{\texttt{#1}}\fi
\expandafter\ifx\csname urlprefix\endcsname\relax\def\urlprefix{URL }\fi
\providecommand{\bibinfo}[2]{#2}
\providecommand{\eprint}[2][]{\url{#2}}

\bibitem[{\citenamefont{Jungwirth et~al.}(2005)\citenamefont{Jungwirth, Wang,
  Masek, Edmonds, K\"{o}nig, Sinova, Polini, Goncharuk, MacDonald, Sawicki
  et~al.}}]{Jungwirthetal:PRB2005}
\bibinfo{author}{\bibfnamefont{T.}~\bibnamefont{Jungwirth}},
  \bibinfo{author}{\bibfnamefont{K.}~\bibnamefont{Wang}},
  \bibinfo{author}{\bibfnamefont{J.}~\bibnamefont{Masek}},
  \bibinfo{author}{\bibfnamefont{K.}~\bibnamefont{Edmonds}},
  \bibinfo{author}{\bibfnamefont{J.}~\bibnamefont{K\"{o}nig}},
  \bibinfo{author}{\bibfnamefont{J.}~\bibnamefont{Sinova}},
  \bibinfo{author}{\bibfnamefont{M.}~\bibnamefont{Polini}},
  \bibinfo{author}{\bibfnamefont{N.}~\bibnamefont{Goncharuk}},
  \bibinfo{author}{\bibfnamefont{A.}~\bibnamefont{MacDonald}},
  \bibinfo{author}{\bibfnamefont{M.}~\bibnamefont{Sawicki}},
  \bibnamefont{et~al.}, \bibinfo{journal}{Phys. Rev. B}
  \textbf{\bibinfo{volume}{72}}, \bibinfo{pages}{165204}
  (\bibinfo{year}{2005}).

\bibitem[{\citenamefont{Edmonds et~al.}(2005)\citenamefont{Edmonds, Farley,
  Johal, van~der Laan, Campion, Gallagher, and Foxon}}]{Edmonds:PRB05}
\bibinfo{author}{\bibfnamefont{K.~W.} \bibnamefont{Edmonds}},
  \bibinfo{author}{\bibfnamefont{N.~R.~S.} \bibnamefont{Farley}},
  \bibinfo{author}{\bibfnamefont{T.~K.} \bibnamefont{Johal}},
  \bibinfo{author}{\bibfnamefont{G.}~\bibnamefont{van~der Laan}},
  \bibinfo{author}{\bibfnamefont{R.~P.} \bibnamefont{Campion}},
  \bibinfo{author}{\bibfnamefont{B.~L.} \bibnamefont{Gallagher}},
  \bibnamefont{and} \bibinfo{author}{\bibfnamefont{C.~T.} \bibnamefont{Foxon}},
  \bibinfo{journal}{Phys. Rev. B} \textbf{\bibinfo{volume}{71}},
  \bibinfo{pages}{064418} (\bibinfo{year}{2005}).

\bibitem[{\citenamefont{Edmonds
  et~al.}(2004{\natexlab{a}})\citenamefont{Edmonds, Campion, Gallagher, Farley,
  Sawicki, Boguslawski, and Dietl}}]{Wangetal:JAP2004}
\bibinfo{author}{\bibfnamefont{K.~W.~K.} \bibnamefont{Edmonds}},
  \bibinfo{author}{\bibfnamefont{R.}~\bibnamefont{Campion}},
  \bibinfo{author}{\bibfnamefont{B.}~\bibnamefont{Gallagher}},
  \bibinfo{author}{\bibfnamefont{N.}~\bibnamefont{Farley}},
  \bibinfo{author}{\bibfnamefont{C.~F.~M.} \bibnamefont{Sawicki}},
  \bibinfo{author}{\bibfnamefont{P.}~\bibnamefont{Boguslawski}},
  \bibnamefont{and} \bibinfo{author}{\bibfnamefont{T.}~\bibnamefont{Dietl}},
  \bibinfo{journal}{J. of Appl. Phys.} \textbf{\bibinfo{volume}{95}},
  \bibinfo{pages}{6512} (\bibinfo{year}{2004}{\natexlab{a}}).

\bibitem[{\citenamefont{Foxon et~al.}(2004)\citenamefont{Foxon, Campion,
  Edmonds, Zhao, Wang, Farley, Staddon, and Gallagher}}]{Foxon:JMatSci04}
\bibinfo{author}{\bibfnamefont{C.~T.} \bibnamefont{Foxon}},
  \bibinfo{author}{\bibfnamefont{R.~P.} \bibnamefont{Campion}},
  \bibinfo{author}{\bibfnamefont{K.~W.} \bibnamefont{Edmonds}},
  \bibinfo{author}{\bibfnamefont{L.}~\bibnamefont{Zhao}},
  \bibinfo{author}{\bibfnamefont{K.}~\bibnamefont{Wang}},
  \bibinfo{author}{\bibfnamefont{N.~R.~S.} \bibnamefont{Farley}},
  \bibinfo{author}{\bibfnamefont{C.~R.} \bibnamefont{Staddon}},
  \bibnamefont{and} \bibinfo{author}{\bibfnamefont{B.~L.}
  \bibnamefont{Gallagher}}, \bibinfo{journal}{J. Mat. Sci.: Mat. El.}
  \textbf{\bibinfo{volume}{15}}, \bibinfo{pages}{727} (\bibinfo{year}{2004}).

\bibitem[{\citenamefont{Edmonds et~al.}(2006)\citenamefont{Edmonds, van~der
  Laan, Freeman, Farley, Johal, Campion, Foxon, Gallagher, and
  Arenholz}}]{Edmondsetal:PRL2006}
\bibinfo{author}{\bibfnamefont{K.}~\bibnamefont{Edmonds}},
  \bibinfo{author}{\bibfnamefont{G.}~\bibnamefont{van~der Laan}},
  \bibinfo{author}{\bibfnamefont{A.}~\bibnamefont{Freeman}},
  \bibinfo{author}{\bibfnamefont{N.}~\bibnamefont{Farley}},
  \bibinfo{author}{\bibfnamefont{T.}~\bibnamefont{Johal}},
  \bibinfo{author}{\bibfnamefont{R.}~\bibnamefont{Campion}},
  \bibinfo{author}{\bibfnamefont{C.}~\bibnamefont{Foxon}},
  \bibinfo{author}{\bibfnamefont{B.}~\bibnamefont{Gallagher}},
  \bibnamefont{and} \bibinfo{author}{\bibfnamefont{E.}~\bibnamefont{Arenholz}},
  \bibinfo{journal}{Phys. Rev. Lett.} \textbf{\bibinfo{volume}{96}},
  \bibinfo{pages}{117207} (\bibinfo{year}{2006}).

\bibitem[{\citenamefont{Burch et~al.}(2006)\citenamefont{Burch, Shrekenhamer,
  Singley, Stephens, Sheu, Kawakami, Schiffer, Samarth, Awschalom, and
  Basov}}]{Burchetal:2006}
\bibinfo{author}{\bibfnamefont{K.}~\bibnamefont{Burch}},
  \bibinfo{author}{\bibfnamefont{D.}~\bibnamefont{Shrekenhamer}},
  \bibinfo{author}{\bibfnamefont{E.}~\bibnamefont{Singley}},
  \bibinfo{author}{\bibfnamefont{J.}~\bibnamefont{Stephens}},
  \bibinfo{author}{\bibfnamefont{B.}~\bibnamefont{Sheu}},
  \bibinfo{author}{\bibfnamefont{R.}~\bibnamefont{Kawakami}},
  \bibinfo{author}{\bibfnamefont{P.}~\bibnamefont{Schiffer}},
  \bibinfo{author}{\bibfnamefont{N.}~\bibnamefont{Samarth}},
  \bibinfo{author}{\bibfnamefont{D.}~\bibnamefont{Awschalom}},
  \bibnamefont{and} \bibinfo{author}{\bibfnamefont{D.}~\bibnamefont{Basov}},
  \bibinfo{journal}{cond-mat/0603851}  (\bibinfo{year}{2006}).

\bibitem[{\citenamefont{Ohno et~al.}(2000)\citenamefont{Ohno, Chilba,
  Matsukura, Omlya, Abe, Dietl, Ohno, and Ohtani}}]{Ohno:Nature00}
\bibinfo{author}{\bibfnamefont{H.}~\bibnamefont{Ohno}},
  \bibinfo{author}{\bibfnamefont{D.}~\bibnamefont{Chilba}},
  \bibinfo{author}{\bibfnamefont{F.}~\bibnamefont{Matsukura}},
  \bibinfo{author}{\bibfnamefont{T.}~\bibnamefont{Omlya}},
  \bibinfo{author}{\bibfnamefont{E.}~\bibnamefont{Abe}},
  \bibinfo{author}{\bibfnamefont{T.}~\bibnamefont{Dietl}},
  \bibinfo{author}{\bibfnamefont{Y.}~\bibnamefont{Ohno}}, \bibnamefont{and}
  \bibinfo{author}{\bibfnamefont{K.}~\bibnamefont{Ohtani}},
  \bibinfo{journal}{Nature} \textbf{\bibinfo{volume}{408}},
  \bibinfo{pages}{944} (\bibinfo{year}{2000}).

\bibitem[{\citenamefont{Wolf et~al.}(2001)\citenamefont{Wolf, Awschalom,
  Buhrman, Daughton, von\ Moln\'ar, Roukes, Chtchelkanova, and
  Treger}}]{Wolf:Science01}
\bibinfo{author}{\bibfnamefont{S.~A.} \bibnamefont{Wolf}},
  \bibinfo{author}{\bibfnamefont{D.~D.} \bibnamefont{Awschalom}},
  \bibinfo{author}{\bibfnamefont{R.~A.} \bibnamefont{Buhrman}},
  \bibinfo{author}{\bibfnamefont{J.~M.} \bibnamefont{Daughton}},
  \bibinfo{author}{\bibfnamefont{S.}~\bibnamefont{von\ Moln\'ar}},
  \bibinfo{author}{\bibfnamefont{M.~L.} \bibnamefont{Roukes}},
  \bibinfo{author}{\bibfnamefont{A.~Y.} \bibnamefont{Chtchelkanova}},
  \bibnamefont{and} \bibinfo{author}{\bibfnamefont{D.~M.}
  \bibnamefont{Treger}}, \bibinfo{journal}{Science}
  \textbf{\bibinfo{volume}{294}}, \bibinfo{pages}{1488} (\bibinfo{year}{2001}).

\bibitem[{\citenamefont{Dietl et~al.}(2000)\citenamefont{Dietl, Ohno,
  Matsukura, Cibert, and Ferrand}}]{Dietl:Science00}
\bibinfo{author}{\bibfnamefont{T.}~\bibnamefont{Dietl}},
  \bibinfo{author}{\bibfnamefont{H.}~\bibnamefont{Ohno}},
  \bibinfo{author}{\bibfnamefont{F.}~\bibnamefont{Matsukura}},
  \bibinfo{author}{\bibfnamefont{J.}~\bibnamefont{Cibert}}, \bibnamefont{and}
  \bibinfo{author}{\bibfnamefont{D.}~\bibnamefont{Ferrand}},
  \bibinfo{journal}{Science} \textbf{\bibinfo{volume}{287}},
  \bibinfo{pages}{1019} (\bibinfo{year}{2000}).

\bibitem[{\citenamefont{Zener}(1950)}]{Zener:PR50}
\bibinfo{author}{\bibfnamefont{C.}~\bibnamefont{Zener}},
  \bibinfo{journal}{Phys. Rev.} \textbf{\bibinfo{volume}{81}},
  \bibinfo{pages}{440} (\bibinfo{year}{1950}).

\bibitem[{\citenamefont{Zener}(1951)}]{Zener:PR51b}
\bibinfo{author}{\bibfnamefont{C.}~\bibnamefont{Zener}},
  \bibinfo{journal}{Phys. Rev.} \textbf{\bibinfo{volume}{83}},
  \bibinfo{pages}{299} (\bibinfo{year}{1951}).

\bibitem[{\citenamefont{Sonoda et~al.}(2002)\citenamefont{Sonoda, Shimizu,
  Sasaki, Yamamoto, and Hori}}]{Sonoda:JCrystGrowth02}
\bibinfo{author}{\bibfnamefont{S.}~\bibnamefont{Sonoda}},
  \bibinfo{author}{\bibfnamefont{S.}~\bibnamefont{Shimizu}},
  \bibinfo{author}{\bibfnamefont{T.}~\bibnamefont{Sasaki}},
  \bibinfo{author}{\bibfnamefont{Y.}~\bibnamefont{Yamamoto}}, \bibnamefont{and}
  \bibinfo{author}{\bibfnamefont{H.}~\bibnamefont{Hori}}, \bibinfo{journal}{J.
  of Crys. Growth} \textbf{\bibinfo{volume}{237-239}}, \bibinfo{pages}{1358}
  (\bibinfo{year}{2002}).

\bibitem[{\citenamefont{Schneider et~al.}(1987)\citenamefont{Schneider,
  Kaufmann, Wilkening, Baeumler, and K\"{o}hl}}]{Schneider}
\bibinfo{author}{\bibfnamefont{J.}~\bibnamefont{Schneider}},
  \bibinfo{author}{\bibfnamefont{U.}~\bibnamefont{Kaufmann}},
  \bibinfo{author}{\bibfnamefont{W.}~\bibnamefont{Wilkening}},
  \bibinfo{author}{\bibfnamefont{M.}~\bibnamefont{Baeumler}}, \bibnamefont{and}
  \bibinfo{author}{\bibfnamefont{F.}~\bibnamefont{K\"{o}hl}},
  \bibinfo{journal}{Phys. Rev. Lett.} \textbf{\bibinfo{volume}{59}},
  \bibinfo{pages}{240} (\bibinfo{year}{1987}).

\bibitem[{\citenamefont{Kreissl et~al.}(1996)\citenamefont{Kreissl, Ulrici,
  El-Metoui, Vasson, Vasson, and Gavaix}}]{Kreissl}
\bibinfo{author}{\bibfnamefont{J.}~\bibnamefont{Kreissl}},
  \bibinfo{author}{\bibfnamefont{W.}~\bibnamefont{Ulrici}},
  \bibinfo{author}{\bibfnamefont{M.}~\bibnamefont{El-Metoui}},
  \bibinfo{author}{\bibfnamefont{A.~M.} \bibnamefont{Vasson}},
  \bibinfo{author}{\bibfnamefont{A.}~\bibnamefont{Vasson}}, \bibnamefont{and}
  \bibinfo{author}{\bibfnamefont{A.}~\bibnamefont{Gavaix}},
  \bibinfo{journal}{Phys. Rev. B} \textbf{\bibinfo{volume}{54}},
  \bibinfo{pages}{10508} (\bibinfo{year}{1996}).

\bibitem[{\citenamefont{Graf et~al.}(2003)\citenamefont{Graf, Gjukic, Hermann,
  Brandt, Stutzmann, and Ambacher}}]{Grafetal:PRB2003}
\bibinfo{author}{\bibfnamefont{T.}~\bibnamefont{Graf}},
  \bibinfo{author}{\bibfnamefont{M.}~\bibnamefont{Gjukic}},
  \bibinfo{author}{\bibfnamefont{M.}~\bibnamefont{Hermann}},
  \bibinfo{author}{\bibfnamefont{M.}~\bibnamefont{Brandt}},
  \bibinfo{author}{\bibfnamefont{M.}~\bibnamefont{Stutzmann}},
  \bibnamefont{and} \bibinfo{author}{\bibfnamefont{O.}~\bibnamefont{Ambacher}},
  \bibinfo{journal}{Phys. Rev. B} \textbf{\bibinfo{volume}{67}},
  \bibinfo{pages}{165215} (\bibinfo{year}{2003}).

\bibitem[{\citenamefont{Korotokov et~al.}(2001)\citenamefont{Korotokov, Gregie,
  Han, and Wessels}}]{Korotokovetal:PhysB2001}
\bibinfo{author}{\bibfnamefont{R.}~\bibnamefont{Korotokov}},
  \bibinfo{author}{\bibfnamefont{J.}~\bibnamefont{Gregie}},
  \bibinfo{author}{\bibfnamefont{B.}~\bibnamefont{Han}}, \bibnamefont{and}
  \bibinfo{author}{\bibfnamefont{B.}~\bibnamefont{Wessels}},
  \bibinfo{journal}{Physica B} \textbf{\bibinfo{volume}{308}},
  \bibinfo{pages}{18} (\bibinfo{year}{2001}).

\bibitem[{\citenamefont{Hwang et~al.}(2005)\citenamefont{Hwang, Ishida,
  Kobayashi, Hirata, Takubo, Mizokawa, Fujimori, Okamoto, Mamiya, Saito
  et~al.}}]{Hwangetal:PRB2005}
\bibinfo{author}{\bibfnamefont{J.}~\bibnamefont{Hwang}},
  \bibinfo{author}{\bibfnamefont{Y.}~\bibnamefont{Ishida}},
  \bibinfo{author}{\bibfnamefont{M.}~\bibnamefont{Kobayashi}},
  \bibinfo{author}{\bibfnamefont{H.}~\bibnamefont{Hirata}},
  \bibinfo{author}{\bibfnamefont{A.}~\bibnamefont{Takubo}},
  \bibinfo{author}{\bibfnamefont{T.}~\bibnamefont{Mizokawa}},
  \bibinfo{author}{\bibfnamefont{A.}~\bibnamefont{Fujimori}},
  \bibinfo{author}{\bibfnamefont{J.}~\bibnamefont{Okamoto}},
  \bibinfo{author}{\bibfnamefont{K.}~\bibnamefont{Mamiya}},
  \bibinfo{author}{\bibfnamefont{Y.}~\bibnamefont{Saito}},
  \bibnamefont{et~al.}, \bibinfo{journal}{Phys. Rev. B}
  \textbf{\bibinfo{volume}{72}}, \bibinfo{pages}{085216}
  (\bibinfo{year}{2005}).

\bibitem[{\citenamefont{Dietl et~al.}(2001)\citenamefont{Dietl, Ohno, and
  Matsukura}}]{Dietl:PRB01}
\bibinfo{author}{\bibfnamefont{T.}~\bibnamefont{Dietl}},
  \bibinfo{author}{\bibfnamefont{H.}~\bibnamefont{Ohno}}, \bibnamefont{and}
  \bibinfo{author}{\bibfnamefont{F.}~\bibnamefont{Matsukura}},
  \bibinfo{journal}{Phys. Rev. B} \textbf{\bibinfo{volume}{63}},
  \bibinfo{pages}{195205} (\bibinfo{year}{2001}).

\bibitem[{\citenamefont{Lee et~al.}(2002)\citenamefont{Lee, Jungwirth, and
  MacDonald}}]{MacDonald:Review02}
\bibinfo{author}{\bibfnamefont{B.}~\bibnamefont{Lee}},
  \bibinfo{author}{\bibfnamefont{T.}~\bibnamefont{Jungwirth}},
  \bibnamefont{and} \bibinfo{author}{\bibfnamefont{A.~H.}
  \bibnamefont{MacDonald}}, \bibinfo{journal}{Semicond. Sci. Technol}
  \textbf{\bibinfo{volume}{17}}, \bibinfo{pages}{393} (\bibinfo{year}{2002}).

\bibitem[{\citenamefont{von Barth and Hedin}(1972)}]{vonBarth:JPhysC72}
\bibinfo{author}{\bibfnamefont{U.}~\bibnamefont{von Barth}} \bibnamefont{and}
  \bibinfo{author}{\bibfnamefont{L.}~\bibnamefont{Hedin}}, \bibinfo{journal}{J.
  Phys. C: Sol. State Phys.} \textbf{\bibinfo{volume}{5}},
  \bibinfo{pages}{1629} (\bibinfo{year}{1972}).

\bibitem[{\citenamefont{Hohenberg and Kohn}(1964)}]{Hohenberg:PR64}
\bibinfo{author}{\bibfnamefont{P.}~\bibnamefont{Hohenberg}} \bibnamefont{and}
  \bibinfo{author}{\bibfnamefont{W.}~\bibnamefont{Kohn}},
  \bibinfo{journal}{Phys. Rev.} \textbf{\bibinfo{volume}{136}},
  \bibinfo{pages}{B864} (\bibinfo{year}{1964}).

\bibitem[{\citenamefont{Kohn and Sham}(1965)}]{Kohn:PR65}
\bibinfo{author}{\bibfnamefont{W.}~\bibnamefont{Kohn}} \bibnamefont{and}
  \bibinfo{author}{\bibfnamefont{L.}~\bibnamefont{Sham}},
  \bibinfo{journal}{Phys. Rev.} \textbf{\bibinfo{volume}{140}},
  \bibinfo{pages}{A1133} (\bibinfo{year}{1965}).

\bibitem[{\citenamefont{Martin}(2004)}]{martin}
\bibinfo{author}{\bibfnamefont{R.~M.} \bibnamefont{Martin}},
  \emph{\bibinfo{title}{Electronic Structure: Basic Theory and Practical
  Methods}} (\bibinfo{publisher}{Cambridge}, \bibinfo{year}{2004}).

\bibitem[{\citenamefont{K\"{u}bler and Eyert}(1994)}]{kubler}
\bibinfo{author}{\bibfnamefont{J.}~\bibnamefont{K\"{u}bler}} \bibnamefont{and}
  \bibinfo{author}{\bibfnamefont{V.}~\bibnamefont{Eyert}}, in
  \emph{\bibinfo{booktitle}{Electronic and Magnetic Properties of Metals and
  Ceramics}}, edited by \bibinfo{editor}{\bibfnamefont{K.~H.~J.}
  \bibnamefont{Bushow}} (\bibinfo{publisher}{VCH}, \bibinfo{address}{Weinheim},
  \bibinfo{year}{1994}), vol.~\bibinfo{volume}{3A}, p.~\bibinfo{pages}{1}.

\bibitem[{\citenamefont{Sato and Katayama-Yoshida}(2001)}]{Sato:JJAP2001}
\bibinfo{author}{\bibfnamefont{K.}~\bibnamefont{Sato}} \bibnamefont{and}
  \bibinfo{author}{\bibfnamefont{H.}~\bibnamefont{Katayama-Yoshida}},
  \bibinfo{journal}{Jpn. J. Appl. Phys.} \textbf{\bibinfo{volume}{40}},
  \bibinfo{pages}{L485} (\bibinfo{year}{2001}).

\bibitem[{\citenamefont{Schulthess and Butler}()}]{Schulthess:JAP01}
\bibinfo{author}{\bibfnamefont{T.~C.} \bibnamefont{Schulthess}}
  \bibnamefont{and} \bibinfo{author}{\bibfnamefont{W.~H.} \bibnamefont{Butler}}
  (????).

\bibitem[{\citenamefont{Sato and Katayama-Yoshida}(2002)}]{Sato:Review02}
\bibinfo{author}{\bibfnamefont{K.}~\bibnamefont{Sato}} \bibnamefont{and}
  \bibinfo{author}{\bibfnamefont{H.}~\bibnamefont{Katayama-Yoshida}},
  \bibinfo{journal}{Semicond. Sci. Technol.} \textbf{\bibinfo{volume}{17}},
  \bibinfo{pages}{367} (\bibinfo{year}{2002}).

\bibitem[{\citenamefont{Mahadevan and
  Zunger}(2004{\natexlab{a}})}]{Mahadevan:PRB04}
\bibinfo{author}{\bibfnamefont{P.}~\bibnamefont{Mahadevan}} \bibnamefont{and}
  \bibinfo{author}{\bibfnamefont{A.}~\bibnamefont{Zunger}},
  \bibinfo{journal}{Phys. Rev. B} \textbf{\bibinfo{volume}{69}},
  \bibinfo{pages}{115211} (\bibinfo{year}{2004}{\natexlab{a}}).

\bibitem[{\citenamefont{Mahadevan and
  Zunger}(2004{\natexlab{b}})}]{Mahadevan:APL04}
\bibinfo{author}{\bibfnamefont{P.}~\bibnamefont{Mahadevan}} \bibnamefont{and}
  \bibinfo{author}{\bibfnamefont{A.}~\bibnamefont{Zunger}},
  \bibinfo{journal}{Appl. Phys. Lett.} \textbf{\bibinfo{volume}{85}},
  \bibinfo{pages}{2860} (\bibinfo{year}{2004}{\natexlab{b}}).

\bibitem[{\citenamefont{Filippetti et~al.}(2005)\citenamefont{Filippetti,
  Spaldin, and Sanvito}}]{Filippetti:JMMM2005}
\bibinfo{author}{\bibfnamefont{A.}~\bibnamefont{Filippetti}},
  \bibinfo{author}{\bibfnamefont{N.}~\bibnamefont{Spaldin}}, \bibnamefont{and}
  \bibinfo{author}{\bibfnamefont{S.}~\bibnamefont{Sanvito}},
  \bibinfo{journal}{J. Magn. Magn. Mater.} \textbf{\bibinfo{volume}{290-291}},
  \bibinfo{pages}{1391} (\bibinfo{year}{2005}).

\bibitem[{\citenamefont{Schulthess et~al.}(2005)\citenamefont{Schulthess,
  Temmerman, Szotek, Butler, and Stocks}}]{Schulthess:NMAT05}
\bibinfo{author}{\bibfnamefont{T.}~\bibnamefont{Schulthess}},
  \bibinfo{author}{\bibfnamefont{W.}~\bibnamefont{Temmerman}},
  \bibinfo{author}{\bibfnamefont{Z.}~\bibnamefont{Szotek}},
  \bibinfo{author}{\bibfnamefont{W.}~\bibnamefont{Butler}}, \bibnamefont{and}
  \bibinfo{author}{\bibfnamefont{G.}~\bibnamefont{Stocks}},
  \bibinfo{journal}{Nature Materials} \textbf{\bibinfo{volume}{4}},
  \bibinfo{pages}{838} (\bibinfo{year}{2005}).

\bibitem[{\citenamefont{Jones and Gunnarsson}(1989)}]{jones}
\bibinfo{author}{\bibfnamefont{R.~O.} \bibnamefont{Jones}} \bibnamefont{and}
  \bibinfo{author}{\bibfnamefont{O.}~\bibnamefont{Gunnarsson}},
  \bibinfo{journal}{Rev. Mod. Phys.} \textbf{\bibinfo{volume}{61}},
  \bibinfo{pages}{689} (\bibinfo{year}{1989}).

\bibitem[{\citenamefont{K\"ubler}(2000)}]{kuebler}
\bibinfo{author}{\bibfnamefont{J.}~\bibnamefont{K\"ubler}},
  \emph{\bibinfo{title}{Theory of Itinerant Electron Magnetism}}, vol.
  \bibinfo{volume}{106} of \emph{\bibinfo{series}{International Series of
  Monographs on Physics}} (\bibinfo{publisher}{Oxford}, \bibinfo{year}{2000}).

\bibitem[{\citenamefont{Aulbur et~al.}(2000)\citenamefont{Aulbur, Jonsson, and
  Wilkins}}]{SSP:2000}
\bibinfo{author}{\bibfnamefont{W.}~\bibnamefont{Aulbur}},
  \bibinfo{author}{\bibfnamefont{L.}~\bibnamefont{Jonsson}}, \bibnamefont{and}
  \bibinfo{author}{\bibfnamefont{J.}~\bibnamefont{Wilkins}},
  \emph{\bibinfo{title}{Quasiparticle calculations in solids}},
  vol.~\bibinfo{volume}{54} of \emph{\bibinfo{series}{SOLID STATE PHYSICS:
  ADVANCES IN RESEARCH AND APPLICATIONS}} (\bibinfo{publisher}{Academic Press
  Inc}, \bibinfo{address}{San Diego, CA 92101-4495 USA}, \bibinfo{year}{2000}).

\bibitem[{\citenamefont{Dreizler and Gross}(1990)}]{DreizlerGross}
\bibinfo{author}{\bibfnamefont{R.}~\bibnamefont{Dreizler}} \bibnamefont{and}
  \bibinfo{author}{\bibfnamefont{E.}~\bibnamefont{Gross}},
  \emph{\bibinfo{title}{Density Functional Theory}}
  (\bibinfo{publisher}{Springer-Verlag}, \bibinfo{address}{Berlin},
  \bibinfo{year}{1990}).

\bibitem[{\citenamefont{Perdew and Zunger}(1981)}]{PZ:SIC}
\bibinfo{author}{\bibfnamefont{J.~P.} \bibnamefont{Perdew}} \bibnamefont{and}
  \bibinfo{author}{\bibfnamefont{A.}~\bibnamefont{Zunger}},
  \bibinfo{journal}{Phys. Rev. B} \textbf{\bibinfo{volume}{23}},
  \bibinfo{pages}{5048} (\bibinfo{year}{1981}).

\bibitem[{\citenamefont{Norman and Koelling}(1984)}]{Norman:PRB84}
\bibinfo{author}{\bibfnamefont{M.~R.} \bibnamefont{Norman}} \bibnamefont{and}
  \bibinfo{author}{\bibfnamefont{D.~D.} \bibnamefont{Koelling}},
  \bibinfo{journal}{Phys. Rev. B} \textbf{\bibinfo{volume}{30}},
  \bibinfo{pages}{5530} (\bibinfo{year}{1984}).

\bibitem[{\citenamefont{Anisimov et~al.}(1991)\citenamefont{Anisimov, Zaanen,
  and Andersen}}]{Anisimov:PRB91}
\bibinfo{author}{\bibfnamefont{V.~I.} \bibnamefont{Anisimov}},
  \bibinfo{author}{\bibfnamefont{J.}~\bibnamefont{Zaanen}}, \bibnamefont{and}
  \bibinfo{author}{\bibfnamefont{O.~K.} \bibnamefont{Andersen}},
  \bibinfo{journal}{Phys. Rev. B} \textbf{\bibinfo{volume}{44}},
  \bibinfo{pages}{943} (\bibinfo{year}{1991}).

\bibitem[{\citenamefont{Sandratskii et~al.}(2004)\citenamefont{Sandratskii,
  Bruno, and Kudrnovsk\'{y}}}]{Sandratskiietal:PRB2004}
\bibinfo{author}{\bibfnamefont{L.}~\bibnamefont{Sandratskii}},
  \bibinfo{author}{\bibfnamefont{P.}~\bibnamefont{Bruno}}, \bibnamefont{and}
  \bibinfo{author}{\bibfnamefont{J.}~\bibnamefont{Kudrnovsk\'{y}}},
  \bibinfo{journal}{Phys. Rev. B} \textbf{\bibinfo{volume}{69}},
  \bibinfo{pages}{195203} (\bibinfo{year}{2004}).

\bibitem[{\citenamefont{Temmerman et~al.}(1998)\citenamefont{Temmerman, Svane,
  Szotek, and Winter}}]{Temmerman:SIC}
\bibinfo{author}{\bibfnamefont{W.~M.} \bibnamefont{Temmerman}},
  \bibinfo{author}{\bibfnamefont{A.}~\bibnamefont{Svane}},
  \bibinfo{author}{\bibfnamefont{Z.}~\bibnamefont{Szotek}}, \bibnamefont{and}
  \bibinfo{author}{\bibfnamefont{H.}~\bibnamefont{Winter}}, in
  \emph{\bibinfo{booktitle}{Electronic Density Functional Theory: Recent
  Progress and New Directions}}, edited by
  \bibinfo{editor}{\bibfnamefont{J.~F.} \bibnamefont{Dobson}},
  \bibinfo{editor}{\bibfnamefont{G.}~\bibnamefont{Vignale}}, \bibnamefont{and}
  \bibinfo{editor}{\bibfnamefont{M.~P.} \bibnamefont{Das}}
  (\bibinfo{publisher}{Plenum}, \bibinfo{address}{New York},
  \bibinfo{year}{1998}), p. \bibinfo{pages}{327}.

\bibitem[{foo()}]{footnote}
\bibinfo{note}{We do not consider interstitial Mn impurities, since they seem
  to reduce the Curie temperature and in experiment can be removed with
  suitable sample preparation.\cite{Gallagher:PRL04}}

\bibitem[{\citenamefont{Kimura et~al.}(1999)\citenamefont{Kimura, Mizokawa,
  Fujimori, Hayashi, and Tanaka}}]{fujimori:PRB1999}
\bibinfo{author}{\bibfnamefont{J.~O.~A.} \bibnamefont{Kimura}},
  \bibinfo{author}{\bibfnamefont{T.}~\bibnamefont{Mizokawa}},
  \bibinfo{author}{\bibfnamefont{A.}~\bibnamefont{Fujimori}},
  \bibinfo{author}{\bibfnamefont{T.}~\bibnamefont{Hayashi}}, \bibnamefont{and}
  \bibinfo{author}{\bibfnamefont{M.}~\bibnamefont{Tanaka}},
  \bibinfo{journal}{Phys. Rev. B} \textbf{\bibinfo{volume}{59}},
  \bibinfo{pages}{R2486} (\bibinfo{year}{1999}).

\bibitem[{\citenamefont{Okabayashi et~al.}(2002)\citenamefont{Okabayashi,
  Mizokawa, Sarma, and Fujimori}}]{fujimori:PRB2002}
\bibinfo{author}{\bibfnamefont{J.}~\bibnamefont{Okabayashi}},
  \bibinfo{author}{\bibfnamefont{T.}~\bibnamefont{Mizokawa}},
  \bibinfo{author}{\bibfnamefont{D.}~\bibnamefont{Sarma}}, \bibnamefont{and}
  \bibinfo{author}{\bibfnamefont{A.}~\bibnamefont{Fujimori}},
  \bibinfo{journal}{Phys. Rev. B} \textbf{\bibinfo{volume}{65}},
  \bibinfo{pages}{161203(R)} (\bibinfo{year}{2002}).

\bibitem[{\citenamefont{Freeman et~al.}(1987)\citenamefont{Freeman, Min, and
  Norman}}]{Freeman:87}
\bibinfo{author}{\bibfnamefont{A.~J.} \bibnamefont{Freeman}},
  \bibinfo{author}{\bibfnamefont{B.~I.} \bibnamefont{Min}}, \bibnamefont{and}
  \bibinfo{author}{\bibfnamefont{M.~R.} \bibnamefont{Norman}}, in
  \emph{\bibinfo{booktitle}{Handbook on the Physics and Chemistry of Rare
  Earths}}, edited by \bibinfo{editor}{\bibfnamefont{K.~A.} \bibnamefont{Jr.}},
  \bibinfo{editor}{\bibfnamefont{L.}~\bibnamefont{Eyring}}, \bibnamefont{and}
  \bibinfo{editor}{\bibfnamefont{S.}~\bibnamefont{H\"ufner}}
  (\bibinfo{publisher}{Elservier}, \bibinfo{year}{1987}),
  vol.~\bibinfo{volume}{10}, pp. \bibinfo{pages}{165--229}.

\bibitem[{\citenamefont{Krieger
  et~al.}(1992{\natexlab{a}})\citenamefont{Krieger, Li, and
  Iafrate}}]{iafrate:PRA92}
\bibinfo{author}{\bibfnamefont{J.}~\bibnamefont{Krieger}},
  \bibinfo{author}{\bibfnamefont{Y.}~\bibnamefont{Li}}, \bibnamefont{and}
  \bibinfo{author}{\bibfnamefont{G.}~\bibnamefont{Iafrate}},
  \bibinfo{journal}{Phys. Rev. A} \textbf{\bibinfo{volume}{45}},
  \bibinfo{pages}{101} (\bibinfo{year}{1992}{\natexlab{a}}).

\bibitem[{\citenamefont{Krieger
  et~al.}(1992{\natexlab{b}})\citenamefont{Krieger, Li, and
  Iafrate}}]{iafrate1:PRA92}
\bibinfo{author}{\bibfnamefont{J.}~\bibnamefont{Krieger}},
  \bibinfo{author}{\bibfnamefont{Y.}~\bibnamefont{Li}}, \bibnamefont{and}
  \bibinfo{author}{\bibfnamefont{G.}~\bibnamefont{Iafrate}},
  \bibinfo{journal}{Phys. Rev. A} \textbf{\bibinfo{volume}{46}},
  \bibinfo{pages}{5453} (\bibinfo{year}{1992}{\natexlab{b}}).

\bibitem[{\citenamefont{Li et~al.}(1993)\citenamefont{Li, Krieger, and
  Iafrate}}]{iafrate:PRA93}
\bibinfo{author}{\bibfnamefont{Y.}~\bibnamefont{Li}},
  \bibinfo{author}{\bibfnamefont{J.}~\bibnamefont{Krieger}}, \bibnamefont{and}
  \bibinfo{author}{\bibfnamefont{G.}~\bibnamefont{Iafrate}},
  \bibinfo{journal}{Phys. Rev. A} \textbf{\bibinfo{volume}{47}},
  \bibinfo{pages}{165} (\bibinfo{year}{1993}).

\bibitem[{\citenamefont{Kotani}(1998)}]{kotani:JPCM98}
\bibinfo{author}{\bibfnamefont{T.}~\bibnamefont{Kotani}}, \bibinfo{journal}{J.
  Phys.: Condens. Matter} \textbf{\bibinfo{volume}{10}}, \bibinfo{pages}{9241}
  (\bibinfo{year}{1998}).

\bibitem[{\citenamefont{Svane et~al.}(2006)\citenamefont{Svane, Christensen,
  Petit, Szotek, and Temmerman}}]{DMSrareearths}
\bibinfo{author}{\bibfnamefont{A.}~\bibnamefont{Svane}},
  \bibinfo{author}{\bibfnamefont{N.}~\bibnamefont{Christensen}},
  \bibinfo{author}{\bibfnamefont{L.}~\bibnamefont{Petit}},
  \bibinfo{author}{\bibfnamefont{Z.}~\bibnamefont{Szotek}}, \bibnamefont{and}
  \bibinfo{author}{\bibfnamefont{W.}~\bibnamefont{Temmerman}},
  \bibinfo{journal}{cond-mat/0603288}  (\bibinfo{year}{2006}).

\bibitem[{\citenamefont{Filippetti and Spaldin}(2003)}]{Filippetti:PRB2003}
\bibinfo{author}{\bibfnamefont{A.}~\bibnamefont{Filippetti}} \bibnamefont{and}
  \bibinfo{author}{\bibfnamefont{N.}~\bibnamefont{Spaldin}},
  \bibinfo{journal}{Phys. Rev. B} \textbf{\bibinfo{volume}{67}},
  \bibinfo{pages}{125109} (\bibinfo{year}{2003}).

\bibitem[{\citenamefont{Shick et~al.}(2004)\citenamefont{Shick, Kudrnovsk\'{y},
  and Drchal}}]{Shicketal:PRB2004}
\bibinfo{author}{\bibfnamefont{A.}~\bibnamefont{Shick}},
  \bibinfo{author}{\bibfnamefont{J.}~\bibnamefont{Kudrnovsk\'{y}}},
  \bibnamefont{and} \bibinfo{author}{\bibfnamefont{V.}~\bibnamefont{Drchal}},
  \bibinfo{journal}{Phys. Rev. B} \textbf{\bibinfo{volume}{69}},
  \bibinfo{pages}{125207} (\bibinfo{year}{2004}).

\bibitem[{\citenamefont{Petit et~al.}(2006)\citenamefont{Petit, Schulthess,
  Svane, Szotek, Temmerman, and Janotti}}]{Petit:PRB2006}
\bibinfo{author}{\bibfnamefont{L.}~\bibnamefont{Petit}},
  \bibinfo{author}{\bibfnamefont{T.}~\bibnamefont{Schulthess}},
  \bibinfo{author}{\bibfnamefont{A.}~\bibnamefont{Svane}},
  \bibinfo{author}{\bibfnamefont{Z.}~\bibnamefont{Szotek}},
  \bibinfo{author}{\bibfnamefont{W.}~\bibnamefont{Temmerman}},
  \bibnamefont{and} \bibinfo{author}{\bibfnamefont{A.}~\bibnamefont{Janotti}},
  \bibinfo{journal}{Phys. Rev. B} \textbf{\bibinfo{volume}{73}},
  \bibinfo{pages}{045107} (\bibinfo{year}{2006}).

\bibitem[{\citenamefont{Ernst et~al.}(2005)\citenamefont{Ernst, Sandratskii,
  Bouhassoune, Henk, and L\"{u}ders}}]{Ernst:PRL2005}
\bibinfo{author}{\bibfnamefont{A.}~\bibnamefont{Ernst}},
  \bibinfo{author}{\bibfnamefont{L.}~\bibnamefont{Sandratskii}},
  \bibinfo{author}{\bibfnamefont{M.}~\bibnamefont{Bouhassoune}},
  \bibinfo{author}{\bibfnamefont{J.}~\bibnamefont{Henk}}, \bibnamefont{and}
  \bibinfo{author}{\bibfnamefont{M.}~\bibnamefont{L\"{u}ders}},
  \bibinfo{journal}{Phys. Rev. Lett.} \textbf{\bibinfo{volume}{95}},
  \bibinfo{pages}{237207} (\bibinfo{year}{2005}).

\bibitem[{\citenamefont{Edmonds
  et~al.}(2004{\natexlab{b}})\citenamefont{Edmonds, Boguslawski, Wang, Campion,
  Novikov, Farley, Gallagher, Foxon, Sawicki, Dietl et~al.}}]{Gallagher:PRL04}
\bibinfo{author}{\bibfnamefont{K.~W.} \bibnamefont{Edmonds}},
  \bibinfo{author}{\bibfnamefont{P.}~\bibnamefont{Boguslawski}},
  \bibinfo{author}{\bibfnamefont{K.~Y.} \bibnamefont{Wang}},
  \bibinfo{author}{\bibfnamefont{R.~P.} \bibnamefont{Campion}},
  \bibinfo{author}{\bibfnamefont{S.~N.} \bibnamefont{Novikov}},
  \bibinfo{author}{\bibfnamefont{N.~R.~S.} \bibnamefont{Farley}},
  \bibinfo{author}{\bibfnamefont{B.~L.} \bibnamefont{Gallagher}},
  \bibinfo{author}{\bibfnamefont{C.~T.} \bibnamefont{Foxon}},
  \bibinfo{author}{\bibfnamefont{M.}~\bibnamefont{Sawicki}},
  \bibinfo{author}{\bibfnamefont{T.}~\bibnamefont{Dietl}},
  \bibnamefont{et~al.}, \bibinfo{journal}{Phys. Rev. Lett.}
  \textbf{\bibinfo{volume}{92}}, \bibinfo{pages}{037201}
  (\bibinfo{year}{2004}{\natexlab{b}}).

\end{thebibliography}


\end{document}